\documentclass[useAMS,usenatbib]{mn2e}
\usepackage[cp866]{inputenc}
\usepackage[english]{babel}
\usepackage{graphicx}
\usepackage{amsmath}
\usepackage{amssymb}
\usepackage{epsf}
\usepackage{bm}

\newcommand{\rate}{\mbox{${\rm erg~cm^{-3}~s^{-1}}$}}
\newcommand{\gcc}{\mbox{${\rm g~cm^{-3}}$}}
\title[Magnetars as cooling neutron stars with internal heating]
{Magnetars as cooling neutron stars with internal heating}
\author[A.~D.~Kaminker et al.]
{ A.~D.~Kaminker$^{1}$,     
          D.~G.~Yakovlev$^{1}$,          
	  A.~Y.~Potekhin$^{1}$, \\
	  {}\\	  
	  {\rm \LARGE N.~Shibazaki$^{2}$,
	  P.~S.~Shternin$^{1}$,
          and
          O.~Y.~Gnedin$^{3}$} \\
	  {}\\
$^{1}$Ioffe Physico-Technical Institute,
Politekhnicheskaya 26, 194021 Saint-Petersburg, Russia\\
$^{2}$Rikkyo University, Tokyo 171-8501, Japan\\
$^{3}$Ohio State University,
Columbus, OH 43210, USA
}
\begin{document}
\date{Accepted 2006 xxxx. Received 2006 xxxx; 
in original form 2006 xxxx}

\pagerange{\pageref{firstpage}--\pageref{lastpage}} \pubyear{2006}

\maketitle
\label{firstpage}

\begin{abstract}
We study thermal structure and evolution of magnetars
as cooling neutron stars with a phenomenological
heat source in a spherical internal layer.
We explore the location of this layer as well as
the heating rate that could explain high observable
thermal luminosities of magnetars
and would be consistent with the energy budget of neutron stars.
We conclude that the heat source should be located
in an outer magnetar's crust, at densities 
$\rho \lesssim 5 \times 10^{11}$~g~cm$^{-3}$,
and should have the heat intensity of $\sim  10^{20}$ \rate.
Otherwise the heat energy is mainly emitted by neutrinos
and cannot warm up the surface. 
\end{abstract}


%%%%%%%%%%%%%%%%%%%%%%%%%%%%%%%%%%%%%%%%%%%%%%%%%%%%%%%%%%%%%%%%%%%%%%%%%%%%%
%**************** Section 1 ******************************
\section{Introduction}
\label{introduction}
%%%%%%%%%%%%%%%%%%%%%%%%%%%%%%%%%%%%%%%%%%%%%%%%%%%%%%%%%%%%%

There is growing evidence that soft gamma repeaters (SGRs)
and anomalous X-ray pulsars (AXPs) belong to the same class
of objects, {\it magnetars}, which are warm, isolated 
slowly rotating neutron stars of age
$t \lesssim 10^5$ yr with unusually strong magnetic fields,
$B \gtrsim 10^{14}$~G (see, e.g., 
\citealt{wt06},
%Woods \& Thompson 2006, 
for a recent review). There have been
attempts to explain the activity of these
sources and the high level of their X-ray emission by
the release of the magnetic energy in their interiors
but a reliable theory is still absent.

In this paper we analyze the thermal evolution
of magnetars as cooling isolated neutron stars.
We do not attempt to construct a self-consistent
theory of the magnetars but instead address the problem
phenomenologically. We show that magnetars
are too hot to be treated as purely cooling neutron stars;
they require some heating source, which we assume
operates in their interiors. Our aim is
to analyze the location and power of the heating
source that are consistent with observed thermal
luminosities of SGRs and AXPs and with the energy budget
of an isolated neutron star.

%%%%%%%%%%%%%%%%%%%%%%%%%%%%%%%%%%%%%%%%%%%%%%%%%%%%%%%%%%%%%%%%
\section{Physics input}
\label{physics}
%%%%%%%%%%%%%%%%%%%%%%%%%%%%%%%%%%%%%%%%%%%%%%%%%%%%%%%%%%%%%%%%

We use our general relativistic 
cooling code (Gnedin, Yakovlev \& Potekhin 2001).
It simulates the thermal evolution of an
initially hot isolated neutron star
taking into account heat outflow via neutrino emission 
from the star and via thermal
conduction within the star, with subsequent thermal photon
emission from the surface. To facilitate
calculations, the star is artificially divided
%(e.g., Gudmundsson, Pethick \& Epstein 1983)
(e.g., \citealt{gpe83})
into a thin outer heat blanketing envelope (extending from the surface
to the layer of density $\rho=\rho_{\rm b}\sim 10^{10}-10^{11}$ 
g~cm$^{-3}$, with the thickness of a few hundred meters), 
and the bulk interior (from $\rho_{\rm b}$ to the center). 
In the bulk interior the code solves the full set of equations of
thermal evolution in the spherically symmetric approximation,
neglecting the effects of magnetic fields on thermal
conduction and neutrino emission.
In the blanketing envelope the code uses the solution of
the stationary thermal conduction problem obtained 
in the approximation of a thin plane-parallel layer for
a dipole magnetic field configuration. This solution
relates temperature $T_{\rm b}$ at the base
of the blanketing envelope ($\rho=\rho_{\rm b}$) to
the effective surface temperature $T_{\rm s}$ 
properly averaged over the neutron star surface
(e.g., \citealt{py01},
%Potekhin \& Yakovlev 2001, 
%Potekhin et al.\ 2003
\citealt{potekhinetal03}). 

In the code
we have mainly used the $T_{\rm b}-T_{\rm s}$ relation
obtained specifically for   
the present study, assuming $\rho_{\rm b}=10^{10}$~g~cm$^{-3}$ 
and the magnetized blanketing envelope made of iron. 
Magnetars
are hot and have large temperature
gradients extending deeply into the heat 
blanketing envelope. Thus
even high magnetic
fields do not dramatically affect the overall thermal
conduction in the envelope and the
$T_{\rm b}-T_{\rm s}$ relation
(as discussed, e.g., by 
%Potekhin et al.\ 2003
\citealt{potekhinetal03}).
Since magnetars are hot inside, these fields should
not greatly affect heat transport in the 
bulk of the star (at $\rho > \rho_{\rm b}$). This 
justifies the approximation of spherically symmetric temperature
distribution at $\rho>\rho_{\rm b}$. At lower temperatures
the electron thermal conductivity in the interior
becomes strongly anisotropic and 
such an approximation can be questionable
\citep{gkp04,gkp06}.
%(Geppert, K\"uker, \& Page 2004, 2006).  

Our standard cooling code includes the effects of
magnetic fields only in the blanketing envelope.
In this respect our present model of the blanketing envelope
with $\rho_{\rm b}=10^{10}$~g~cm$^{-3}$ seems less
adequate than the previous model with 
$\rho_{\rm b}=4 \times 10^{11}$~g~cm$^{-3}$ 
%(Potekhin et al.\ 2003). 
\citep{potekhinetal03}.
However, the latter model,
by construction, neglects neutrino emission from the
blanketing envelope. Because the neutrino emission from the
layers with $\rho \gtrsim 3 \times 10^{10}$~g~cm$^{-3}$
in hot magnetars is important (Section \ref{results}), the 
blanketing envelope
with $\rho_{\rm b}=10^{10}$~g~cm$^{-3}$ is expected to be more appropriate
for the present investigation.
That is why we have performed calculations
with this latter model but have done
additional tests using the model 
with $\rho_{\rm b}=4\times 10^{11}$~g~cm$^{-3}$ 
(Section \ref{tests}). 

We have also included, in a phenomenological
manner, the effects of
magnetic fields 
on the thermal evolution of the bulk stellar interior.
Most importantly, we have introduced a
heat source located within a spherical layer
in the interior (it can be 
associated with the magnetic field, Section \ref{nature}). 
The heating rate $H$ [\rate] has been
taken in the form
\begin{equation}
     H=H_0\,\Theta(\rho_1,\rho_2)\,\exp(-t/\tau),
\label{Q}     
\end{equation}
where $H_0$ is the maximum heat intensity,
$\Theta(\rho_1,\rho_2)$ is a step-like function
($\Theta\approx1$ within the density interval
$\rho_1<\rho<\rho_2$; $\Theta \approx 0$ outside
this interval, with a sharp but continuous transitions
at the interval boundaries), $t$ is the star's age,
and $\tau$ is the life time of the heating source. 
A specific form of $H$ is not important for 
our main conclusions.
We do not specify the nature of this
source. We set $\tau=5 \times 10^4$
years to explain high thermal states
of all observed SGRs and AXPs (Section
\ref{results}). We do not consider longer $\tau$
which would require higher energy budget
(while the budget is already severely restricted even for 
$\tau=5 \times 10^4$ years). We treat $H_0$, $\rho_1$ and
$\rho_2$ as free parameters with the aim to
understand what the intensity and
the location of the heat source should be 
in order to be consistent
with observations and with the energy budget of
an isolated neutron star. 

It is instructive to introduce the
total heat power $W^\infty$ [erg~s$^{-1}$],
redshifted for a distant observer,
\begin{equation}
   W^\infty(t)= \int {\rm d}V\,{\rm e}^{2\Phi}\, H,
\label{W}
\end{equation}
where ${\rm d}V$ is the proper volume element,
and $\Phi$ is the redshift metric function.

In the neutron star core we use the
equation of state of dense matter
constructed by 
%Akmal, Pandharipande \& Ravenhall (1998) 
\citet{apr98}
(model Argon V18 + $\delta v$ + UIX$^*$);
it is currently
considered the most elaborated equation of state 
of neutron star matter.
Specifically, we employ 
a convenient parameterization 
of this equation of state proposed by 
%Heiselberg \& Hjorth-Jensen (1999) 
\citet{hhj99}
and described as APR~III by
%Gusakov et al.\ (2005)
\citet{gusakovetal05}. 
%The equation of state is sufficiently
%stiff. 
According to this equation of state,
neutron star cores consist of neutrons, protons,
electrons, and muons.  The maximum (gravitational) neutron star mass
is $M=1.929\,M_\odot$. The powerful direct Urca process of
neutrino emission 
%(Lattimer et al.\ 1991)
\citep{lpph91}
is allowed only in the central kernels
of massive neutron stars with $M > 1.685\,M_\odot$ (at densities
$\rho> 1.275 \times 10^{15}$ g~cm$^{-3}$).

We use neutron star models with two masses,
$M=1.4\,M_\odot$ and $M=1.9\,M_\odot$.
The $1.4\,M_\odot$ model is an example of a star with
standard (not too strong) neutrino emission in the core
(provided by the modified Urca process in a non-superfluid
star). In this case the (circumferential) stellar radius is
$R=12.27$~km, and the central density is 
$\rho_{\rm c}=9.280 \times 10^{14}$~g~cm$^{-3}$.
The $1.9\,M_\odot$ model ($R=10.95$~km, $\rho_{\rm c}=
2.050 \times 10^{15}$~g~cm$^{-3}$) is an example of a star
whose neutrino emission is enhanced by the direct Urca process
in the inner core.  

We have updated the thermal
conductivity of electrons and muons in the stellar 
core by new results (P.~S.~Shternin \& 
D.~G.~Yakovlev 2006, in preparation) which take into account 
the Landau damping of transverse plasmons in the interaction
of electrons and muons with surrounding charged particles
(following the results for quark plasma obtained by
%Heiselberg \& Pethick 1993). 
\citealt{hp93}).
This update has not 
noticeably affected our cooling scenarios.
 
%%%%%%%%%%%%%%%%%%%%%%%%%%%%%%%%%%%%%%%%%%%%%%%%%%%%%%%%%%%%%%%%%%%%%
\section{Results}
\label{results}
%%%%%%%%%%%%%%%%%%%%%%%%%%%%%%%%%%%%%%%%%%%%%%%%%%%%%%%%%%%%%%%%%%%%%

%%%%%%%%%%%%%%%%%%%%%%%%%%%%%%%%%%%%%%%%%%%%%%%%%%%%%%%%%%%%%%%%%%%%%
\begin{figure}%[t]
%\centering
\epsfysize=80mm
\epsffile{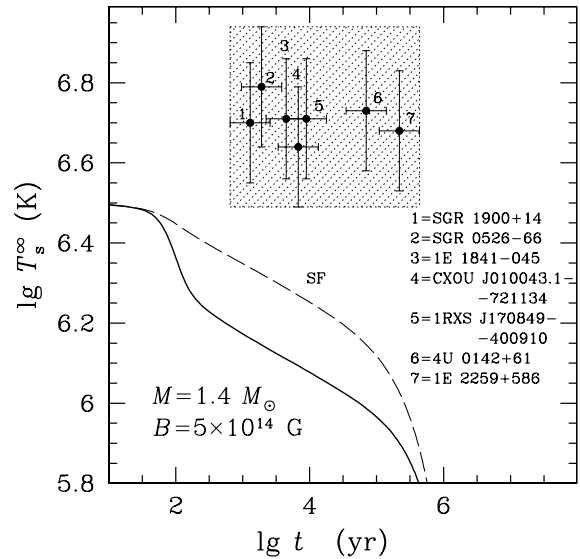}
\caption{
Observational data on the blackbody
surface temperatures $T_{\rm s}^\infty$ of seven magnetars.
The shaded rectangle is the ``magnetar box''.
The data are compared to the theoretical
cooling curves of the
1.4\,$M_\odot$ neutron star with $B=5 \times 10^{14}$~G 
and no internal heating, either without superfluidity (the solid
line) or with strong proton superfluidity in the core
(the dashed line SF).
}
\label{fig1}
\end{figure}
%%%%%%%%%%%%%%%%%%%%%%%%%%%%%%%%%%%%%%%%%%%%%%%%%%%%%%%%%%%%%%%%%%%%%%%%%

%%%%%%%%%%%%%%%%%%%%%%%%%%%%%%%%%%%%%%%%%%%%%%%%%%%%%%%%%%%%%%%%%%%%%%
\renewcommand{\arraystretch}{1.2}
\begin{table*}%[!t]   % "*" ignores the twocolumn-format if adopted
\caption[]{Four positions of the heating layer, and the
heating power $W^\infty$ for the 1.4\,$M_\odot$ star
with $H_0=3\times 10^{20}$ \rate\ and $t=1$ kyr}
\label{tab:heat}
\begin{center}
\begin{tabular}{ c  c  c  c }
\hline
\hline
No. & $\rho_1$ (\gcc) & $ \rho_2$ (\gcc) & $W^\infty$ (erg s$^{-1}$) \\
\hline
\hline
I    & $3 \times 10^{10}$ & $10^{11}$ & $4.0 \times 10^{37}$ \\
II   & $10^{12}$ & $3 \times 10^{12}$ & $1.9 \times 10^{37}$  \\
III  & $3 \times 10^{13}$ & $10^{14}$ & $1.1 \times 10^{38}$ \\
IV   & $3 \times 10^{13}$ & $9 \times 10^{14}$ & $1.1 \times 10^{39}$ \\
\hline
\end{tabular}
\end{center}
\small{
%\begin{flushleft}
%$^{a)}$ Inferred using a hydrogen atmosphere model\\
%\end{flushleft}
}
\end{table*}
%%%%%%%%%%%%%%%%%%%%%%%%%%%%%%%%%%%%%%%%%%%%%%%%%%%%%%%%%%%%%%%%

For the observational basis we take
seven sources (two SGRs and five AXPs indicated in Figure \ref{fig1}).
The estimates of their spindown ages $t$, surface magnetic 
fields $B$ and the blackbody surface
temperatures $T_{\rm s}^\infty$ 
(redshifted for a distant observer)
are taken from Tables 14.1 and
14.2 of the review paper by 
%Woods \& Thompson (2006).
\citet{wt06} and from the paper by \citet{mcgarryetal05}.
The data are from the original publications
of %Kulkarni et al.\ (2003) 
\citet{kulkarnietal03}
(SGR 0526--66);  
%Woods et al.\ (2001,2002) 
\citet{woodsetal01,woodsetal02}
(SGR 1900+14);
%Gotthelf et al.\ (2004)  
\citet{gotthelfetal02}
and 
%Morii et al.\ (2003) 
\citet{moriietal03}
(1E 1841--045);
\citet{mcgarryetal05} (CXOU J010043.1--721134);
%Gavriil \& Kaspi (2002) 
\citet{gk02}
and
%Rea et al.\ (2003) 
\citet{reaetal03}
(1RXS J170849--400910); 
%Patel et al.\ (2003) 
\citet{gk02}
%Gavriil \& Kaspi (2002)
and
\citet{pateletal03} 
(4U 0142+61);
\citet{gk02}
and
%Woods et al.\ (2004) 
\citet{woodsetal04} 
%Gavriil \& Kaspi (2002) 
(1E 2259+586).
In the absence of $T_{\rm s}^\infty$ estimates for SGR 1627--41 
we do not include this SGR in our analysis.
We have also excluded SGR 1806--20 and several AXPs whose
thermal emission component and characteristic age seem less certain. 
The radiation from the selected sources has the pulsed fraction
$\lesssim 20\%$. This indicates that the thermal radiation can be
emitted from a substantial part of the surface. 

The blackbody surface
temperatures $T_{\rm s}^\infty$ 
%(redshifted for a distant observer) 
of the selected sources are plotted
in Figure \ref{fig1} versus spindown ages $t$. 
%Woods \& Thompson (2006) 
\citet{wt06}
present these data without
formal errors, which are actually large.
We introduce, somewhat arbitrarily, the $30\%$ uncertainties
into the values of $T_{\rm s}^\infty$ 
and the uncertainties by a factor of 2
into the values of $t$. The data are too 
uncertain and our cooling models
are too simplified to explain every
source by its own cooling model.
Instead, we try to explain the existence
of magnetars as cooling neutron stars that belong
to the ``magnetar box'', the shaded rectangle
in Figure \ref{fig1} (attributed to
an average persistent thermal emission 
from magnetars, excluding bursting states). Our results will be sufficiently
insensitive to the neutron star mass, and we will
mainly use the neutron star model with $M=1.4\,M_\odot$
(unless the contrary is indicated).
Similarly, the results will be not too
sensitive to superfluid state of stellar
interiors and we will mostly neglect the effects
of superfluidity of nucleons in the stellar crust and
core. To be specific, we mainly assume the dipole
magnetic field in the blanketing envelope with
$B=5\times 10^{14}$~G at the magnetic poles.
Some variations of $B$ will not affect our
principal conclusions (Section \ref{tests}).

In Figure \ref{fig1} we show the theoretical cooling
curves $T_{\rm s}^\infty (t)$ for the 1.4\,$M_\odot$
isolated magnetized neutron star without any internal
heating. The solid line refers to a nonsuperfluid
neutron star while the dashed line SF is for
strong proton superfluidity in the stellar core.
This superfluidity strongly 
suppresses neutrino emission in the core and
noticeably increases $T_{\rm s}^\infty$
at the neutrino cooling stage 
%It gives nearly the
%maximum $T_{\rm s}^\infty(t)$ that an ordinary middle-aged cooling
%neutron star can have without reheating
%(see, e.g., 
%%Yakovlev \& Pethick 2004
(e.g., \citealt{yp04}).
Let us stress that the surface temperature of
these stars is highly nonuniform, with the magnetic poles
being much hotter than the equator. In the figures
we plot the average surface temperature (e.g., \citealt{potekhinetal03}). 
Clearly, the magnetars are much hotter
than ordinary cooling neutron stars.
The observations of ordinary neutron stars can be
explained by the cooling theory without any
reheating (e.g., 
%Yakovlev \& Pethick 2004,
\citealt{yp04};
%Page, Geppert \& Weber 2006), 
\citealt{pgw06}),
while the observations
of magnetars suggest that the magnetars have
additional heat sources. We assume further that
these sources are located inside magnetars.
According to alternative models, powerful energy sources 
can be available in
magnetar magnetospheres 
\citep{tb05,bt06}.
%(Thompson \& Beloborodov 2005).

\begin{figure*}%[t]
%\centering
\epsfysize=80mm
\epsffile{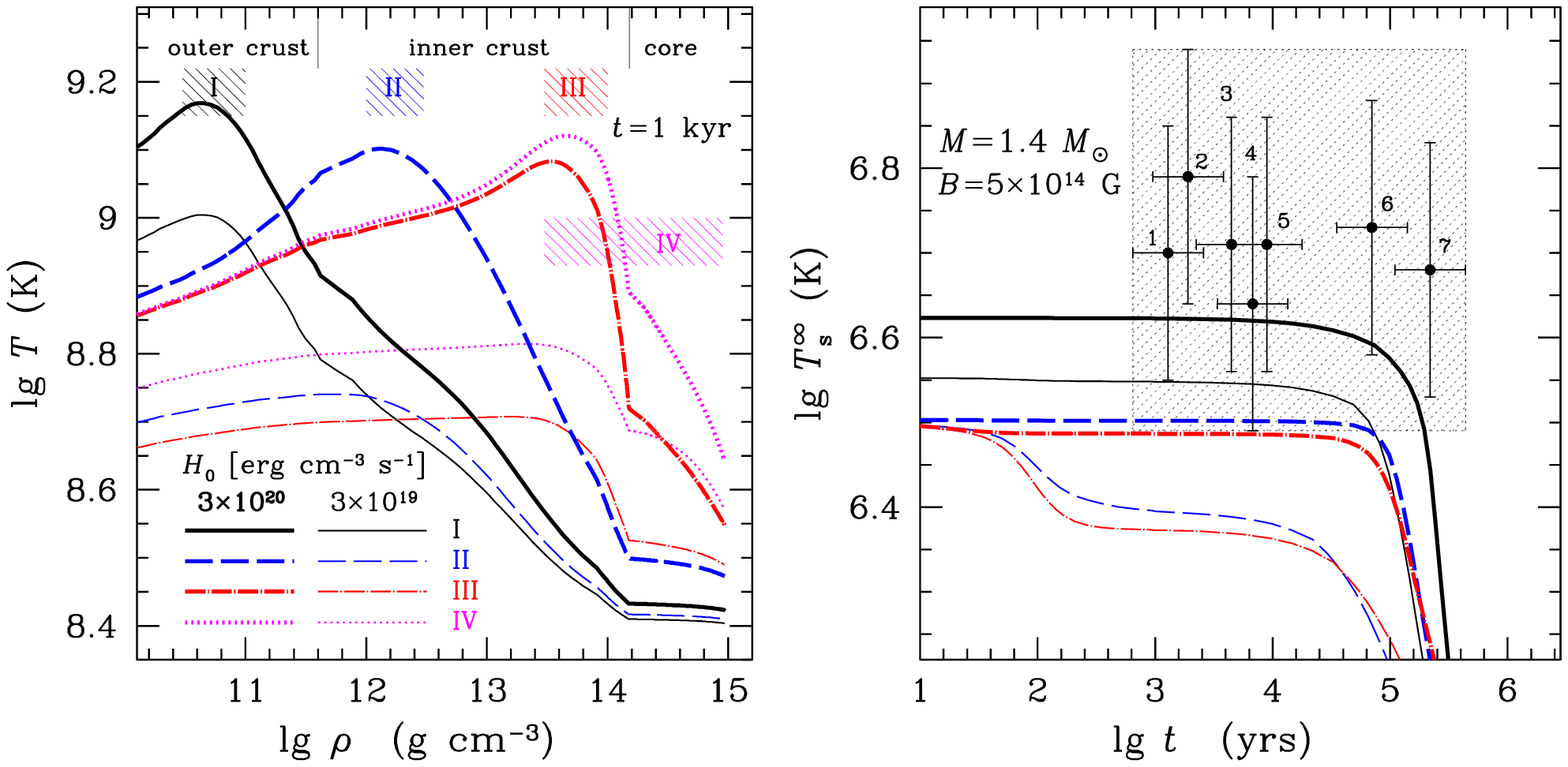}
\caption{ 
{\it Left:} Temperature profiles within the 1.4$\,M_\odot$ neutron
star of age $t=1000$ years with four different 
positions I--IV of the heating
layer (given in Table \ref{tab:heat} 
and indicated by hatched rectangles) and 
two levels of the heat intensity $H_0 = 3\times 10^{19}$
and $3\times10^{20}$ \rate. The magnetic field
is $B=5 \times 10^{14}$~G. {\it Right:} Cooling curves
for these models compared with the observations.
}
\label{fig2}
\end{figure*}

\begin{figure}%[t]
%\centering
\epsfysize=80mm
\epsffile{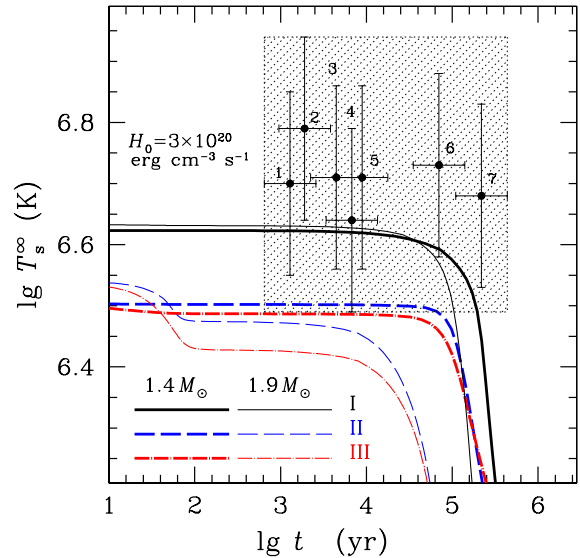}
\caption{
Same as in the right panel of Figure \ref{fig2}
for the three positions of the heating layer,
one value of the heat intensity and two
neutron star masses, 1.4 and 1.9\,$M_\odot$.
}
\label{fig3}
\end{figure}

\begin{figure*}%[t]
%\centering
\epsfysize=80mm
\epsffile{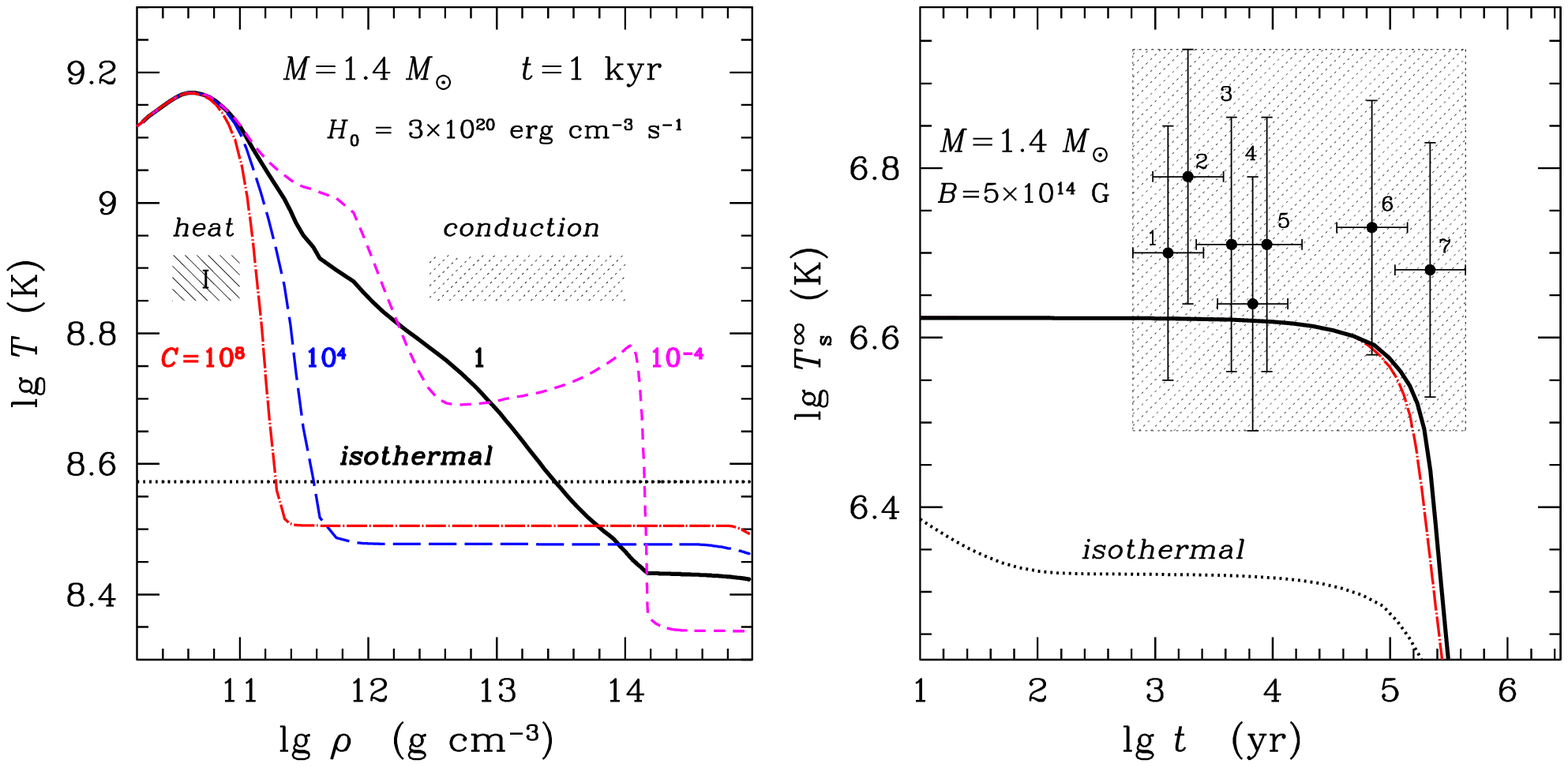}
\caption{
The effect of the thermal conductivity in the
inner crust on the thermal evolution of the 1.4\,$M_\odot$
magnetar with the heating layer I and
the heat intensity 
$H_0=3 \times 10^{20}$ \rate. {\it Left:} The temperature
profiles within the magnetar at $t=$1000 years. 
The hatched rectangles show the positions of
the heating layer and the layer, where the
thermal conductivity was modified. {\it Right:}
The cooling curves. The thick solid lines are the same
as in Figure \ref{fig2}. Thinner 
long-dash,  dot-and-dash, and short-dash lines are
for the star with the thermal conductivity modified
by a factor of $C=10^4$, $10^8$, and $10^{-4}$, respectively, in
the density range from $3 \times 10^{12}$ to $10^{14}$ \gcc. 
The dotted line marked {\it isothermal} is for an infinite
thermal conductivity in the star bulk 
($\rho>10^{10}$ \gcc).
}
\label{fig4}
\end{figure*}

\begin{figure*}%[t]
%\centering
\epsfysize=80mm
\epsffile{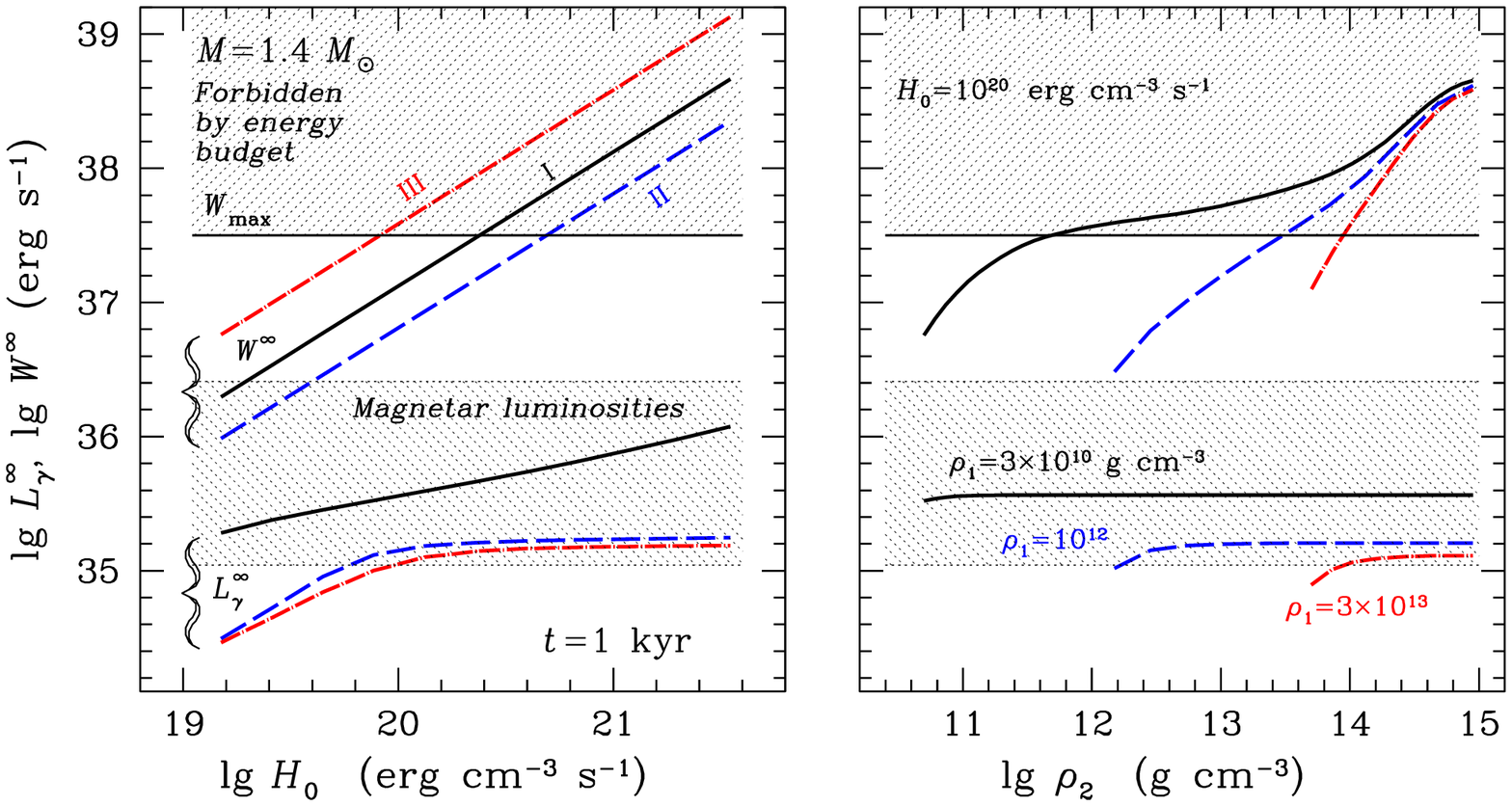}
\caption{
The total heat power $W^\infty$ 
(higher curves) and
the surface photon luminosity $L_\gamma^\infty$
(lower curves) versus parameters of the heating layer
compared to the values of $L_\gamma^\infty$
from the ``magnetar box'' (the lower shaded strip)
and to the values of $W^\infty$ forbidden by energy budget
(the upper shading) for the 1.4\,$M_\odot$ 
neutron star of age 1000 years.
{\it Left:} Three fixed positions of the heating layer
of variable heat intensity $H_0$. {\it Right:} Three fixed
minimum densities $\rho_1$ of the heating layer,
the fixed heat intensity $H_0=10^{20}$ \rate\ and 
variable maximum density $\rho_2$.
}
\label{fig5}
\end{figure*}

We introduce the internal
sources in a phenomenological way 
described in Section \ref{physics}. All results
presented below (Figures \ref{fig2}--\ref{fig5})
are obtained including the internal heating 
in accordance with Eq.\ (\ref{Q}).
They indicate that the magnetars are hot
inside, with the temperature $T\sim 10^9$~K
(or even higher) in the crust, at $\rho \gtrsim
3 \times 10^{10}$~\gcc. Such stars are  
very strong sources of neutrino emission, which is
vitally important for the magnetar thermal structure
and evolution.

Our simulations of cooling neutron stars with
a powerful internal heating show that after a short
initial relaxation ($t \lesssim 10$ years) the 
star reaches a quasi-stationary state, which is fully
determined by the heating source. The energy
is mainly carried away by neutrinos, but some 
fraction is transported by thermal conduction to the
surface and radiated away by photons. The interior
of these cooling neutron stars is highly non-isothermal.
The hotter layers are those where the heat is released
and where the neutrino emission is not too strong; 
these layers are located in the crust. 
The approximation of isothermal interior, which is
excellent for ordinary middle-aged neutron stars,
cannot be used while studying the cooling of magnetars.

%%%%%%%%%%%%%%%%%%%%%%%%%%%%%%%%%%%%%%%%%%%%%%%%%%%%%%%%%%%%%%%%%%%%%
%%%%%%%%%%%%%%%%%%%%%%%%%%%%%%%%%%%%%%%%%%%%%%%%%%%%%%%%%%%%%%%%%%%%%%%%%

The left panel of Figure \ref{fig2} shows the temperature distribution
inside the 1.4\,$M_\odot$ star of age $t=1000$ years.
This distribution remains the same during the
entire magnetar stage ($t \lesssim \tau$, see Eq.~(\ref{Q}))
and we have chosen $t=1000$ years just as an example.
We have considered four locations of the heat
layer, $\rho_1- \rho_2$, summarized in Table
\ref{tab:heat}: (I) $3\times 10^{10}-10^{11}$ \gcc\
(in the outer crust, just below the heat blanketing envelope),
(II) $10^{12}- 3\times 10^{12}$ \gcc\ (at the top
of the inner crust), (III) $3 \times 10^{13}-10^{14}$
\gcc\ (at the bottom of the inner crust), and
(IV) $3 \times 10^{13}-9 \times 10^{14}$ \gcc\ (at the bottom of the inner
crust and in the entire core).
These locations are marked by hatched rectangles.
The density ranges, which are appropriate to the
outer crust, the inner crust and the core, are indicated
in the upper part of the figure. 
Let us remind the reader that the outer crust has a thickness of
a few hundred meters and a mass of $\sim 10^{-5}\,M_\odot$,
the inner crust can be as thick as 1 km and its mass
is $\sim 10^{-2}\,M_\odot$, while the core is large 
(radius $\sim 10$ km) and contains $\sim 99\%$ of
the stellar mass. Therefore, the heating layers I, II, and III
are relatively thin, while the heating layer IV is wide and
includes most of the stellar volume.
We have allowed
our heat sources to have two different intensities,
$H_0=3\times10^{19}$ and $3\times10^{20}$ \rate . 
For illustration, in Table \ref{tab:heat} we present
also the heating power $W^\infty$ calculated from
Eq.\ (\ref{W}) for the four layers in the 1.4\,$M_\odot$ star of age
$t=1000$ years at $H_0=3 \times 10^{20}$ erg s$^{-1}$.

In all the cases the neutron star core appears to be much colder 
than the crust because the core quickly cools down via
strong neutrino emission (via the modified Urca process
for the conditions in Figure \ref{fig2}).
Placing the heat sources far from the heat
blanketing envelope is an inefficient way to
maintain warm surface; the heating layer can be
hot, but the energy is radiated away by neutrinos
and does not flow to the surface. For the deep
heating layers (cases II, III, or IV), the heat intensity
$H_0=3 \times 10^{19}$ \rate\ is clearly insufficient to
warm the surface to the magnetar level. 
The higher intensity $3 \times 10^{20}$ \rate\
helps but is still less efficient in these layers
than in the layer I, which is 
close to the heat blanketing envelope. For the latter $H_0$
the heating of the crust bottom (case III) and the heating
of the entire core (case IV) lead to the same surface
temperature of the star. Thus, the most
efficient way to warm the surface is to place
the heating layer in the outer crust,
near the heat blanketing envelope.
This conclusion is further supported by Figures 
\ref{fig3}--\ref{fig5}. 

The right panel of Figure \ref{fig2} shows cooling curves
of the 1.4\,$M_\odot$ stars for the same models of the heating layer
as in the left panel (with one exception --- we do not show the
cooling curves for the case IV, for simplicity).
The cooling curves of the star of age $t\lesssim 5 \times
10^{4}$ years are almost horizontal, indicating
that these stars maintain their high thermal state
owing to the internal heating. Non-horizontal
initial parts ($t \lesssim 100$ years) of two curves,
which correspond to deep and non-intense
heating, show the residual initial relaxation to 
quasi-stationary thermal states.
One can see that the only way to explain
the sources from the ``magnetar box''
is to place the source in the outer crust and
assume $H_0 \sim 10^{20}$ \rate.
As the heating is exponentially switched off at $t \gtrsim 5 \times 10^4$
years, following Eq.\ (\ref{Q}), the surface temperature
drops down accordingly. The magnetar stage is over
and the star transforms into the
ordinary cooling neutron star 
\citep{potekhinetal03}
%(Potekhin et al.\ 2003)
which cools mainly via the surface photon emission.

%%%%%%%%%%%%%%%%%%%%%%%%%%%%%%%%%%%%%%%%%%%%%%%%%%%%%%%%%%%%%%%%%%%%%
%%%%%%%%%%%%%%%%%%%%%%%%%%%%%%%%%%%%%%%%%%%%%%%%%%%%%%%%%%%%%%%%%%%%%%%%%

Figure \ref{fig3} compares
cooling curves of neutron stars of two
masses, 1.4\,$M_\odot$ and $1.9\,M_\odot$,
for the same three locations I--III of the heating layer
and for one heat intensity $H_0=3 \times 10^{20}$ \rate . 
The thick lines are for the 1.4\,$M_\odot$ star;
they are the same as in the right panel of Figure \ref{fig2}. 
The thin lines
are the respective curves for the 1.9\,$M_\odot$ star,
which is a very efficient neutrino emitter
because of the direct Urca process in its core
(Section \ref{physics}). If the heating layer is 
in the location II or III (in the inner crust),
the enhanced neutrino emission
from the core of the massive star carries away a substantial
amount of heat and decreases the surface
temperature of the star. If, however, the heating
layer is placed in the outer crust (case I),
direct Urca in the core has
almost no effect on the surface temperature.
In this case
the surface temperature  
is almost insensitive to the physical conditions
in the stellar core and in 
the inner crust, in particular, to the neutrino
emission mechanisms and superfluid state of matter.
In other words, surface layers are thermally decoupled
from the deep interior. A similar situation occurs in ordinary
young and hot cooling stars in the initial cooling
stage, before internal thermal relaxation 
%(Lattimer et al.\ 1994, Gnedin et al.\ 2001). 
(\citealt{lattimeretal94},
 \citealt{gyp01}).
This justifies our neglect of 
the effects of superfluidity 
in the calculations (although these effects
can be vitally important for ordinary middle-aged 
cooling neutron stars;
e.g., 
%Yakovlev \& Pethick 2004, Page et al.\ 2006).
\citealt{yp04}, \citealt{pgw06}).

%%%%%%%%%%%%%%%%%%%%%%%%%%%%%%%%%%%%%%%%%%%%%%%%%%%%%%%%%%%%%%%%%%%%%
%%%%%%%%%%%%%%%%%%%%%%%%%%%%%%%%%%%%%%%%%%%%%%%%%%%%%%%%%%%%%%%%%%%%%%%%%

Figure \ref{fig4} demonstrates sensitivity of
the results to the values of the thermal conductivity
in the inner neutron star crust. 
It shows temperature profiles in the 1.4\,$M_\odot$
star of age 1000 years and the cooling curves of this star
for one location of the heating layer (case I)
and one heat intensity
$H_0=3 \times 10^{20}$ \rate.
The thick lines are the same as in Figure \ref{fig2}.
They are calculated with our standard cooling code
assuming only electron thermal conductivity in the
crust 
%(Gnedin et al.\ 2001). 
\citep{gyp01}.
However, in the
inner crust thermal energy can also be transported 
by free neutrons and this transport can be very
efficient, especially if neutrons are in a superfluid
state. The effect may be similar to that in superfluid
$^4$He, where no temperature gradients can be created
in laboratory experiments because they are immediately
smeared out by responding convective flows
(e.g., 
%Tilley \& Tilley 1990). 
\citealt{tt90}).
To simulate this effect
we have artificially introduced the layer of 
high thermal conductivity in the inner neutron star
crust, in the density range from $3 \times 10^{12}$ \gcc\ to 
$10^{14}$ \gcc. In this layer
we enhanced the thermal conductivity by a factor
of $C=10^4$ or $10^8$. The enhancement changes
dramatically the temperature profiles in the inner
crust, making it much cooler and almost isothermal.
We have also made one test run by reducing the
thermal conductivity in the indicated layer by a factor
of $10^4$. This corresponds to $C=10^{-4}$ and
mainly increases the
temperature in the inner crust. However, all these strong
changes of the temperature profiles in the inner crust have almost
no effect on the surface temperature and the cooling curves.
It is another manifestation of the thermal decoupling
of the surface layers from the inner parts of the star.  
In addition, we have simulated
the cooling of the star (the dotted curves) in the approximation of
infinite thermal conductivity everywhere
in the star bulk ($\rho > \rho_{\rm b}=10^{10}$ \gcc).
In this case the heat energy is instantly spread over the star
bulk, which makes the stellar surface much cooler than
in the case of finite conduction.

%%%%%%%%%%%%%%%%%%%%%%%%%%%%%%%%%%%%%%%%%%%%%%%%%%%%%%%%%%%%%%%%%%%%%
%%%%%%%%%%%%%%%%%%%%%%%%%%%%%%%%%%%%%%%%%%%%%%%%%%%%%%%%%%%%%%%%%%%%%%%%%

Figure \ref{fig5} shows the integrated heating rate 
$W^\infty$, given by Eq.\ (\ref{W}) (three upper lines
on each panel), and the
photon thermal surface luminosity of the star
$L_\gamma^\infty$ (redshifted for a distant
observer, three lower lines on each panel) 
as a function of parameters
of the heating layers. The results
are presented for the 1.4\,$M_\odot$ star of age 1000 years.

In the left panel we select three locations of the
heating layer (I, II, III) and vary the heat intensity
$H_0$. One can clearly see that only the heating layer~I
can produce $L_\gamma^\infty \gtrsim 3 \times 10^{35}$
erg~s$^{-1}$, typical for magnetars. Moreover, the surface
luminosity increases with $H_0$ 
much slower than the heating rate. For $H_0 \gtrsim 10^{20}$ \rate\
and the layers II and III,
the luminosity is seen to {\it saturate}. This means that
{\it pumping additional heat energy into the heating layer
does not affect $L_\gamma^\infty$}. The efficiency of
converting the input heat into the surface
radiation ($L_\gamma^\infty/W^\infty$) is generally
small. The highest efficiency is achieved if we heat the
outer crust (the layer I) with low intensity.  

In the right panel of Figure \ref{fig5} we present
$L_\gamma^\infty$ and $W^\infty$ as a function of the
maximum density $\rho_2$ of the heating layer, for 
one heat intensity $H_0=10^{20}$ \rate\ and three
fixed minimum densities of this layer ($\rho_1=3 \times 10^{10}$,
$10^{12}$, and $3 \times 10^{13}$ \gcc).
One can observe the saturation of $L_\gamma^\infty$
with increasing $\rho_2$. If $\rho_2$ is large and
the heating layer is extended into the core, the heat
energy is huge. However, as long as $\rho_1$ is far from
the surface, this huge energy is almost fully carried
away by neutrinos and does not heat the surface.
In this case the efficiency of heat conversion into the surface
emission is very low. 

In Figure \ref{fig5}
we compare qualitatively the calculated values of $L_\gamma^\infty$
with the thermal
surface luminosities from the ``magnetar box''
(the lower shaded strip, estimated using 
accepted values of $T_{\rm s}^\infty$ and the
1.4\,$M_\odot$ neutron star model). The heating should be sufficiently
intense to raise $L_\gamma^\infty$ to these magnetar values.
It is the first requirement to explain the
observations of the magnetars. The second requirement stems
from the energy budget of an isolated neutron star.
The heat energy is assumed to be taken from some source
which pumps the energy $W^\infty$ into the heating layer during
magnetar's life ($\tau\sim 5 \times 10^4$ years in Eq.\ (\ref{Q})).
Naturally, the total available energy $E_{\rm tot}$ is restricted.
At least it should be much smaller than 
$\sim 5 \times 10^{53}$~erg, the gravitational
energy of the neutron star. We assume
that the maximum energy of the internal heating is
$E_{\rm max} \sim 10^{50}$ erg (which is the 
magnetic energy of the star with 
the magnetic field $B=3\times 10^{16}$~G in the core).
%or the thermal energy
%of the newborn star with the internal temperature of
%$\sim 10^{10}$~K). 
Then the maximum energy generation rate
is $W_{\rm max}\sim E_{\rm max}/\tau \sim 3 \times 10^{37}$
erg~s$^{-1}$, which is plotted by the upper horizontal
solid line in Figure \ref{fig5}. The upper shaded space
above this line is thus prohibited by the energy budget.

A successful interpretation of
magnetars as cooling neutron stars
requires $L_\gamma^\infty$ to be sufficiently high
to reach the ``magnetar box'' and $W^\infty$ to be sufficiently
low to avoid the prohibited region.  
These conditions are fulfilled for the heating
layer located in the outer stellar crust and
the heat intensities $H_0$  between $3\times 10^{19}$
and $3\times 10^{20}$ \rate\ (higher value preferred). 
A typical efficiency of heat
conversion into the surface emission under these
conditions is $L_\gamma^\infty/W^\infty \sim 10^{-2}$.

%%%%%%%%%%%%%%%%%%%%%%%%%%%%%%%%%%%%%%%%%%%%%
\section{Discussion}
\label{discussion}
%%%%%%%%%%%%%%%%%%%%%%%%%%%%%%%%%%%%%%%%%%%%%

%%%%%%%%%%%%%%%%%%%%%%%%%%%%%%%%%%%%%%%%%%%%%
\subsection{Testing the results}
\label{tests}
%%%%%%%%%%%%%%%%%%%%%%%%%%%%%%%%%%%%%%%%%%%%%

We have performed a number of additional tests.
In particular, we have studied sensitivity
of our results to the values of the neutrino emissivity
in the neutron star core and crust. We find that
variations of the neutrino emissivity in the inner crust
and the core of the star that undergoes 
intense heating can strongly affect 
the temperature profiles in the stellar interior
but have almost no effect on the surface temperature.
This is another example of the thermal
decoupling of the outer crust and deep interiors.
On the other hand, we find that
$L_\gamma^\infty$ is sensitive
to the neutrino emission in the outer 
crust. This emission in a hot crust is mainly generated by
plasmon decay and electron-nucleus bremsstrahlung
mechanisms (see, e.g., 
%Yakovlev et al.\ 2001). 
\citealt{yakovlevetal01}).
If magnetars
have very strong magnetic fields in their outer crusts,
these fields can greatly modify the plasmon decay
neutrino process. Such modifications have not been
studied in detail but can be important for the magnetar physics. 

We have made some test runs taking into account
the effects of superfluidity. In particular, we
have included superfluidity of free neutrons
in the inner crust and the associated
neutrino emission via Cooper pairing of neutrons.
This neutrino emission affects the temperature
profiles in the inner crust but has almost no effect
on the surface temperature. 

We have also varied the maximum density $\rho_{\rm b}$ of the
heat blanketing layer, shifting it from the present position
$\rho_{\rm b}=10^{10}$
\gcc\ (Section \ref{physics}) to
$\rho_{\rm b}=4 \times 10^{11}$ \gcc, as 
in our previous model 
%(Potekhin et al.\ 2003).
\citep{potekhinetal03}.
In addition, we have varied the magnetic field strength $B$
in the blanketing envelope from
$B=5\times 10^{13}$~G to $5\times 10^{15}$~G.
The cooling curves and the internal temperature profiles
are sensitive to these variations but do not violate
our principal conclusions. It would be desirable to
reconsider the cooling with a more careful treatment
of heat transport and neutrino emission
in a magnetized outer crust. 

%%%%%%%%%%%%%%%%%%%%%%%%%%%%%%%%%%%%%%%%%%%%%%%%%%%%%%%%%%
\subsection{The nature of internal heating}
\label{nature}
%%%%%%%%%%%%%%%%%%%%%%%%%%%%%%%%%%%%%%%%%%%%%%%%%%%%%%%%%%

The development of a specific theoretical model
of the internal heating is outside the scope of this 
paper (existing models are summarized, e.g.,
in review papers by 
%Woods \& Thompson 2006
%and Heyl 2006). 
\citealt{wt06} and \citealt{heyl06}).
Nevertheless, our results place
stringent constraints on the possible models.
The main requirement is that the total heat energy should be huge,
$E_{\rm tot} \sim 10^{49}-10^{50}$ erg, and should be released
during $10^{4}-10^{5}$ years in the outer neutron
star crust. This does not necessarily require
that the energy be stored in the outer crust but
that it be effectively transformed into heat there.

Our results agree with the widespread point of view that
the magnetars can be powered neither by their rotation,
nor by accretion, nor by thermal energy in a cooling star, 
nor by the strain energy accumulated
in the crust. All these energy sources contain much
less than $10^{49}$ erg.
% For instance, even if we
% adopt the maximum strain of the crust ($\sim 0.01$), the
% strain energy would be $\sim 4 \times 10^{44}$ erg, much
% less than required. 

Nevertheless, the required energy can be accumulated in the
magnetic field if, for example, the star possesses the
field $B \sim (1-3)\times10^{16}$~G in its core.
%(or higher field $B \sim 10^{17}$~G in the crust).
The evolution of this magnetic field can be accompanied by
a strong energy release in the outer crust, where
the electric conductivity is especially low and
the field undergoes the strongest Ohmic dissipation.
An actual structure of the magnetic field  
in the star can be complicated. In particular, the
magnetic configuration in the crust can strongly deviate
from a magnetic dipole.

One cannot exclude that thermal radiation
of magnetars is emitted from a smaller part of the neutron star
surface (e.g., from hot spots near magnetic poles), 
implying lower total heat energies. Nevertheless,
observable thermal X-ray luminosities of magnetars
$\sim 10^{34}-10^{36}$ erg~s$^{-1}$ (e.g.,
\citealt{mereghettietal02,kg04,wt06}) are consistent
with the interval of thermal luminosities in Figure \ref{fig5}
calculated assuming the emission from the entire surface. 
Radiation from the sources included
in our ``magnetar box'' usually demonstrates low pulsed fraction
($\lesssim 20\%$) which also indicates that this radiation
can be emitted from the entire surface 
(although the pulsed fraction can be lowered by the
gravitational lensing effect).

Let us recall also alternative theories of magnetars.
They suggest that the main energy release occurs in
the magnetar's magnetosphere (e.g., \citealt{bt06}) and
the radiation spectrum is formed there as
a result of comptonization and reprocession of the
quasi-thermal spectrum.

%%%%%%%%%%%%%%%%%%%%%%%%%%%%%%%%%%%%%%%%%%%%%%%%%%%%%%%%%%
\section{Conclusions}
\label{conclusions}
%%%%%%%%%%%%%%%%%%%%%%%%%%%%%%%%%%%%%%%%%%%%%%%%%%%%%%%%%%

We have modeled thermal states and
thermal evolution of magnetars (SGRs and AXPs)
with the aim to explain high surface
temperatures of these neutron stars and
their energy budget. Our main conclusions
are as follows.

(1) It is impossible to explain high thermal
states of magnetars as cooling neutron stars
without assuming that they
undergo powerful heating. We have developed the
idea that the heat source operates in the interior of magnetars.

(2) The heat source can be located in a thin 
layer at densities $\rho \lesssim 5 \times 10^{11}$~\gcc\ in the
outer magnetar crust, with the
heat intensity ranged from $\sim
3 \times 10^{19}$ to $3 \times 10^{20}$ \rate. 
The source cannot be located deeper 
in the interior because
the heat energy would be radiated away by neutrinos; it
would be unable to warm up the surface.
This deeper heating is extremely inefficient
and inconsistent with the energy budget of neutron stars.
Pumping huge heat energy into the deeper layers 
would not increase the surface temperature.

(3) Heating of the outer crust
produces a strongly nonuniform temperature distribution
within the star. The temperature in the heating layer
exceeds $10^9$~K, while the bottom of the crust and
the stellar core remain much colder. The thermal state
of the heat layer and of the surface is almost
independent of physical parameters of deeper
layers (such as the equation of state, neutrino
emission, heat transport, superfluidity
of baryons), which means thermal decoupling of
the outer crust from the inner layers. 
The total energy released in the heat
layer during magnetar's life ($\sim 10^4-10^5$ years)
cannot be lower than $10^{49}-10^{50}$ erg; maximum
1\% of this energy can be spent to heat the surface.

The present calculations can be improved 
by a more careful treatment of heat conduction
in a magnetized plasma of the outer crust, at
$\rho>10^{10}$ \gcc, as well as by considering different
magnetic fields and the presence of light
(accreted) elements in the surface layers of magnetars.
We intend to study these effects in our next publication. 

The nature and the physical model of internal heating 
are still not clear; they should be elaborated in the future.
In any case, one should bear in mind that the heat
energy released within the magnetars should be at least two
orders of magnitude higher than the photon thermal energy
emitted through their surface, and the energy release should
take place in the outer stellar crust. These 
model-independent conclusions are consequences of the 
well-known principle that any hot, dense matter in
stellar objects is a strong source of neutrino emission.

\section*{Acknowledgments}
We are grateful to V.~D.\ Pal'shin, G.~G.\ Pavlov,
Yu.~A.\ Shibanov, V.~A.~Urpin,
and J.\ Ventura for fruitful discussions and 
critical remarks. This work was partly supported by 
the Russian Foundation for Basic Research
(grants 05-02-16245, 05-02-22003),
by the Federal Agency for Science and Innovations
(grant NSh 9879.2006.2), by the
Rikkyo University Invitee Research Associate Program,
and by the Dynasty Foundation.

%%%%%%%%%%%%%%%%%%%%%%%%%%%%%%%%%%%%%%%%%%%%%%%%%%%%%%%%%%%

\end{document}